\documentclass[a4paper]{article}
\usepackage[T1]{fontenc}

\usepackage{hyperref}
\hypersetup{  colorlinks=false,
  breaklinks=true}
  
\usepackage[american]{babel}

\usepackage[margin=1in]{geometry} %

\usepackage[sort&compress,numbers,square]{natbib}
\renewcommand*{\bibfont}{\footnotesize}%

\usepackage[dvipsnames]{xcolor}

\usepackage{graphicx, color}

\usepackage{amsmath,amsfonts,amssymb}
\usepackage{mathtools}

\usepackage{lastpage}
\usepackage{fancyhdr}
\newcommand{\dd}{\mathrm{d}}

\newcommand{\edit}[1]{#1}

\begin{document}

\title{\textsf{Stochastic switching of delayed feedback suppresses  oscillations in genetic regulatory systems}}

\author{%
Bhargav R. Karamched$^{1,2,3}$, Christopher E. Miles$^{4}$\\
\url{bkaramched@fsu.edu}, \,\url{chris.miles@uci.edu}}

\date{\footnotesize $^{1}$Department of Mathematics, Florida State University, Tallahassee, FL 32304\\
$^{2}$Institute of Molecular Biophysics, Florida State University, Tallahassee, FL 32304\\
$^{3}$Program in Neuroscience, Florida State University, Tallahassee, FL 32304\\
$^{4}$Department of Mathematics, University of California, Irvine, CA 92697}

\maketitle

\begin{abstract}
\noindent Delays and stochasticity have both served as crucially valuable ingredients in mathematical descriptions of control, physical, and biological systems. In this work, we investigate how explicitly dynamical stochasticity in delays modulates the effect of delayed feedback. To do so, we consider a hybrid model where stochastic delays evolve by a continuous-time Markov chain, and between switching events, the system of interest evolves via a deterministic delay equation. Our main contribution is the calculation of an effective delay equation in the fast switching limit. This effective equation maintains the influence of all subsystem delays and cannot be replaced with a single effective delay. To illustrate the relevance of this calculation, we investigate a simple model of stochastically switching delayed feedback motivated by gene regulation. We show that sufficiently fast switching between two oscillatory subsystems can yield stable dynamics. \vspace{.1in}  
\end{abstract}

\tableofcontents 

\section{Introduction}

Mathematical models with delays have served profoundly useful in capturing the behavior of complex systems in  biology \cite{mackey1977oscillation,bressloff2015frequency,karamched2015delayed,karamched2021delay,bai2022closed,karamched2022delay,godin2022space}, networks \cite{krtolica1994stability}, and control \cite{nilsson1998stochastic}. One notable example is delayed negative feedback control of genetic networks, especially transcriptional feedback \cite{ribeiro2010stochastic,gupta2013transcriptional,gupta2014modeling}. . These systems, \edit{such as in nF-$\kappa$B  \cite{hoffmann2002ikappab} or p53 \cite{geva2006oscillations}},  share the canonical setup of some molecule  that autoinhibits its own production with delayed feedback arising from several molecular events that must occur in sequence \cite{vogel1994rna}. The behavior of these models for a fixed $\tau$ is rich but well-understood: the amount of delay in feedback crucially determines stability or instability (often to oscillations) in the system \cite{mackey1977oscillation,xiong2023physiological}.

At the scale of molecular machinery associated with genetic regulation, dynamics are also known to be richly stochastic \cite{mcadams1999sa,elowitz2002stochastic}, with inherent noise in the counts of individual molecules, but also disparate timescales of promotion or inhibitory factors binding and unbinding \cite{bressloff2017stochastic}. Putting these two pieces together, there is a natural interest in understanding the emergent dynamics in systems with delayed feedback and stochasticity in biological  \cite{schlicht2008delay,galla2009intrinsic,yuan2009stochastic,josic2011stochastic,zavala2014delays,fatehi2020new,cao2020analytical,potoyan2014dephasing,zambrano2015interplay} and physical systems \cite{ohira2000delayed,klinshov2020mode}.

There have been many insightful investigations into systems with \textit{distributed delay}. In these models, a delay is continuously drawn from some probability distribution thought to arise from stochasticity or uncertainty in the delay \cite{bernard2009distributed,kyrychko2018enhancing}. Distributed delays in negative feedback can produce interesting behaviors, including bimodality \cite{chiu2022distributed} and surprising stability changes \cite{gupta2013transcriptional}. However, it remains unclear to what extent it is possible to relate distributed delays to more explicit dynamical descriptions of the underlying process that governs them. Others have considered discrete-time models or so-called \textit{semi-discretized} where the delays themselves switch at discrete times \cite{sadeghpour2014stability,gomez2014exact,gomez2016stability,sadeghpour2018can,sadeghpour2019stability}. Such models may be appropriate for control systems but unrealistic for biological systems where the switching is driven by stochastically timed chemical events \cite{bressloff2014stochastic}. Instead, we return to very early models where the delays themselves follow a continuous-time Markov process \cite{lidskii1965stability,kats1967stability,kolmanovsky2000stochastic}. While there are rigorous works investigating the long-time stability of such models \cite{mao2000stochastic,liu2006stability,tosato2020multi,xiao2021stability}, these arguments are primarily based on Lyapounoff functions that are challenging to find for any specific model.

In this work, we investigate the dynamics of a model that stochastically switches between delays at exponential rates. That is, we consider a stochastic hybrid delay system. The delays evolve via a continuous-time Markov chain, and in between these transitions, the system evolves by a deterministic delay equation. Our primary contribution is the derivation of an effective quasi-steady-state delay system in the limit of fast stochastic switching. Perhaps surprisingly, we find that non-linear systems do not converge to one with a single effective delay but retain the effects of all delays in the original subsystems. Using this computation, we explore the behavior of classical model of delayed negative feedback \cite{bernard2006modelling,stricker2008fast,longo2013dual,gupta2013transcriptional}. With stochastic switching, we show that switching between two oscillatory subsystems can stabilize the system. Altogether, our results add clarity and intrigue to the picture of how stochastic switching and delays intertwine in biological systems, especially those  containing negative feedback as seen in genetic regulation. 

\section{Simple Model of Delayed Negative Feedback}
We begin with a simple model of delayed negative feedback with a single fixed delay $\tau$ and review the role delay plays in destabilizing the fixed point of a dynamical system. We are far from the first to consider such a model or its variants \cite{stricker2008fast,longo2013dual, gupta2013transcriptional,gupta2014modeling}, but we include a brief investigation here for the sake of self-containment of our work. Thereafter we demonstrate that allowing for stochastic switching between distinct delays stabilizes the fixed point, even when all delays involved are past the Hopf bifurcation point of the non-switching system.  

\begin{figure}
    \centering
    \includegraphics[width=.95\textwidth]{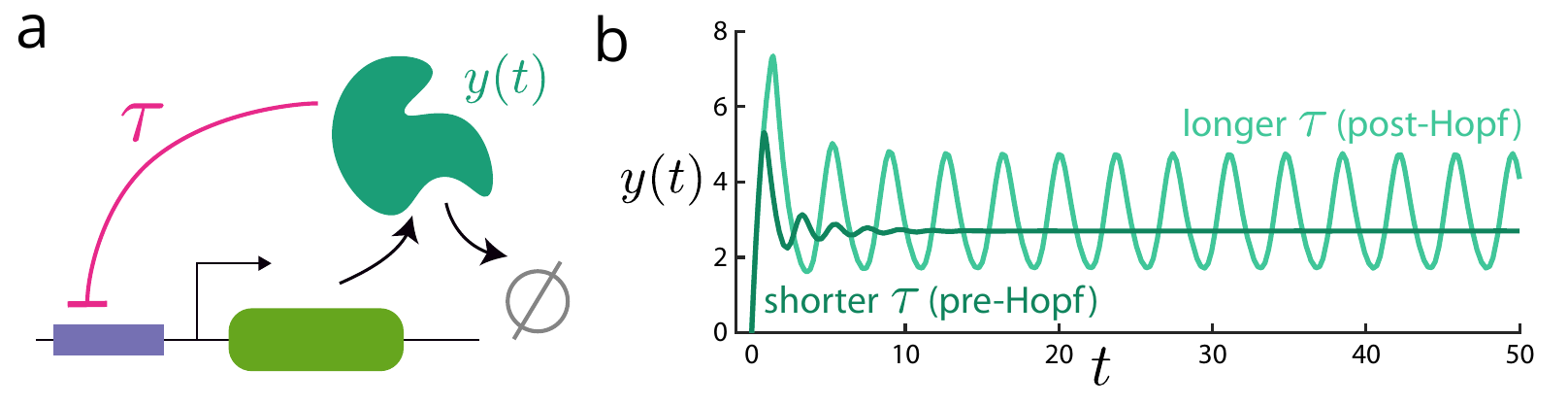}
    \caption{Simple model of delayed negative feedback. \textbf{a}: Schematic of biological motivation for Eq.~\eqref{DNF1}: a protein leads to inhibiting its own production with delay $\tau$, and also degrades. \textbf{b}: Two solutions of Eq.~\eqref{DNF1} for different values of $\tau$. For long delays, oscillations occur, and for short, steady state is achieved. The threshold Hopf bifurcation between these behaviors is described by Eq.~\eqref{hopf1}. }
    \label{fig:DNFintro}
\end{figure}

Let $y(t)$ be a scalar field evolving according to 
\begin{equation}\label{DNF1}
\frac{\dd y}{\dd t} = I - \gamma y - wf(y(t-\tau)).
\end{equation}
Here, $y(t)$ could represent the concentration of a protein that is constitutively produced at a rate $I$ and inhibits its own production. The first order rate constant $\gamma$ describes the natural degradation rate of the substance $y$. We take the weight $w$ to describe the strength of the autoinhibition based on some form of Michaelis-Menten kinetics. Therefore, $f$ can be any function that is monotonically increasing and finite at infinity.  For concreteness, we take
$$
f(y) = \frac{y^n}{K^n + y^n}.
$$
Eq.~\eqref{DNF1} describes a simple model for delayed negative feedback---the substance being produced inhibits its own production following a nonzero but finite time delay $\tau$.
\subsection{Hopf Bifurcation}\label{hopf}
Setting the time derivative equal to zero in Eq.~\eqref{DNF1} yields an equilibrium solution $y^*$ satisfying
\begin{equation}
\label{equilibrium}
\gamma y^* + wf(y^*) = I.
\end{equation}
To understand how $\tau$ destabilizes the equilibrium, we perform a linear stability analysis. Linearizing Eq.~\eqref{DNF1} around $y^*$ yields
$$
\frac{\dd u}{\dd t} = -\gamma u - wf'(y^*)u(t-\tau)
$$
where $u(t) \equiv y(t) - y^*$. This has the solution $u(t) = Ke^{\lambda t}$ with $\lambda$ determined from the eigenvalue equation
\begin{equation}\label{evalue}
\lambda = -\gamma - wf'(y^*)e^{-\lambda \tau}
\end{equation}
In accordance with standard analysis of delay differential
equations, we determine the necessary conditions for the
emergence of a time-periodic solution via a Hopf bifurcation by setting $\lambda = i\omega$ in Eq.~\eqref{evalue}. Equating real and imaginary parts gives the following conditions for a Hopf bifurcation:
\begin{align}\label{hopf1}
\begin{split}
\omega & = wf'(y^*) \sin{(\omega\tau)}\\
-\gamma & = wf'(y^*) \cos{(\omega \tau)}\\
-\frac{\omega}{\gamma} &= \tan{(\omega \tau)}
\end{split}
\end{align}
It is clear that these conditions cannot be satisfied in the absence of delay ($\tau = 0$). Indeed, setting $\tau = 0$ renders Eq.~\eqref{DNF1} a one-dimensional flow, which cannot have oscillations. On the other hand, for $\tau > 0$, there exist ($\omega_c$, $\tau_c$) satisfying conditions \eqref{hopf1}. When $\tau$ is increased past $\tau_c$, a pair of complex conjugate eigenvalues crosses the imaginary axis. Although this is not sufficient to guarantee the emergence
of a stable periodic solution via a supercritical Hopf bifurcation for $\tau > \tau_c$, the existence of stable oscillations beyond
the Hopf bifurcation point can be verified numerically, with examples seen in Fig.~\ref{fig:DNFintro}. In the simulations shown, the parameters chosen are $I=10$, $K=9.5$, $\gamma=1$, $n=4$, and $w=9.5$. These will be used elsewhere throughout the manuscript unless noted otherwise. For these parameter values, $\tau_c \approx 0.9$. The two delays in Fig.~\ref{fig:DNFintro} correspond to values of $\tau$ above and below this bifurcation, $\tau = 0.6$ and $\tau=1.2$ respectively.

Hence, we have established the crucial role delay plays in the destabilization of equilibria corresponding to delayed negative feedback systems.  We next demonstrate the curious result that allowing $\tau$ to randomly switch between two values---both past $\tau_c$---results in the stabilization of the equilibrium.

\subsection{Stochastic Switching \edit{of Delays}}
In biological and biophysical models, delays often manifest as a coarse-grained description of several processes which, cumulatively, require time. Hence, many details underlying biophysical processes are overlooked in fixed-delay systems. To capture finer dynamics, biophysical models incorporate delays that vary according to a probability distribution or evolve via some prescribed dynamics. \edit{We emphasize the contrast of this model setup to those such as in references \citenum{zambrano2015interplay,cao2020analytical}, where stochastic switching of the system occurs (due to inhibitor binding and unbinding) but the delay, if included, remains fixed.} 
   
Here we implement stochasticity in delay equations by taking the delay to evolve in time according to a continuous-time discrete Markov process. Explicitly, consider a general autonomous delay differential equation of the form
\begin{equation}\label{dde1}
\frac{\dd y}{\dd t} = G(y(t), y(t-\tau(t))).
\end{equation}
Here $y(t)$ is the same scalar field and $\tau(t)$ varies in time stochastically between $N \in \mathbb{N}$ states. Let $\tau_i$ denote the delay corresponding to the $i$th state so that $\tau \in \{\tau_1, \tau_2,...,\tau_N\}$ at any given time $t$, and let $Q(\tau_i,t)$ denote the probability that $\tau = \tau_i$ at time $t$. The dynamics of $\tau$ are then completely characterized by the master equation
\begin{equation}\label{master1}
\frac{dQ}{dt}(\tau_i, t) = \sum_{j = 1 \\ j\neq i}^N \mathcal{W}_{j \to i}Q(\tau_j,t) - \mathcal{W}_{i\to j}Q(\tau_i,t),
\end{equation}
where $\mathcal{W}_{i\to j}$ denotes the propensity of the transition $\tau_i \to \tau_j$. Equations~\eqref{dde1} and \eqref{master1} together form a so-called \textit{stochastic hybrid system}---a system wherein the state of the system evolves stochastically, but within each state evolves deterministically. Such systems are also called as a \textit{piecewise deterministic Markov process} (PDMP)~\cite{bressloff2013stochastic,bressloff2014stochastic,bressloff2017stochastic}.

\subsection{Stochastic Delayed Negative Feedback} \label{sdnf}

\begin{figure}
    \centering
    \includegraphics[width=.95\textwidth]{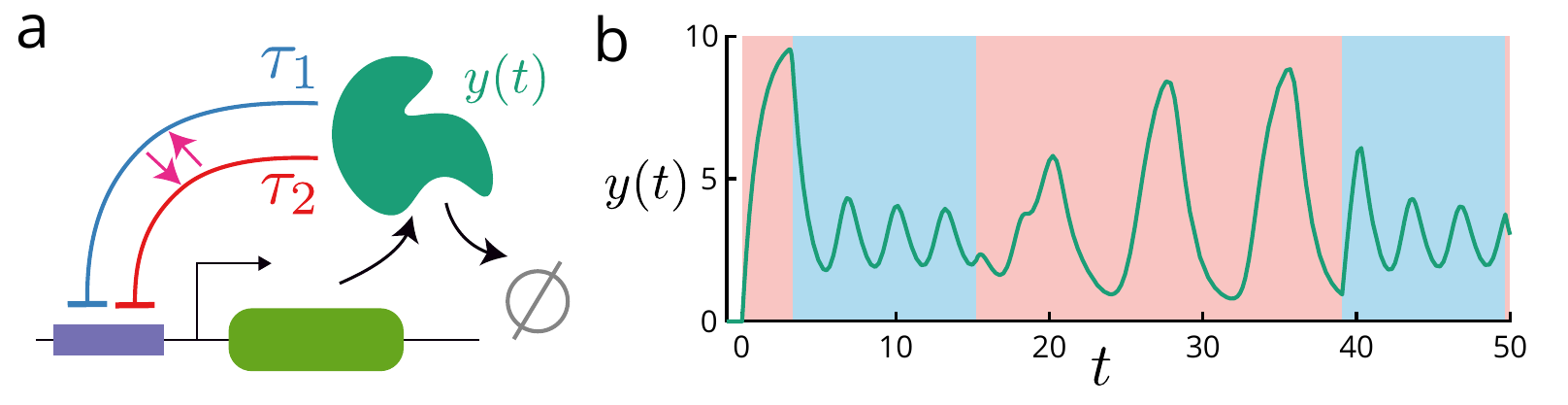}
    \caption{Delayed negative feedback model with stochastic switching. \textbf{a}: Schematic for the biological motivation of the model. A protein with concentration $y(t)$ inhibits its own production with delay that now stochastically switches  between $\tau_1,\tau_2$ at exponentially distributed times. \textbf{b}: Illustrative  simulation of the system, here with relatively slow switching $\varepsilon=10$. In the red regions, the delay state is $\tau_1=3$ and in the blue, $\tau_2=1$. }
    \label{fig:model}
\end{figure}

\begin{figure}
    \centering
    \includegraphics[width=0.75\textwidth]{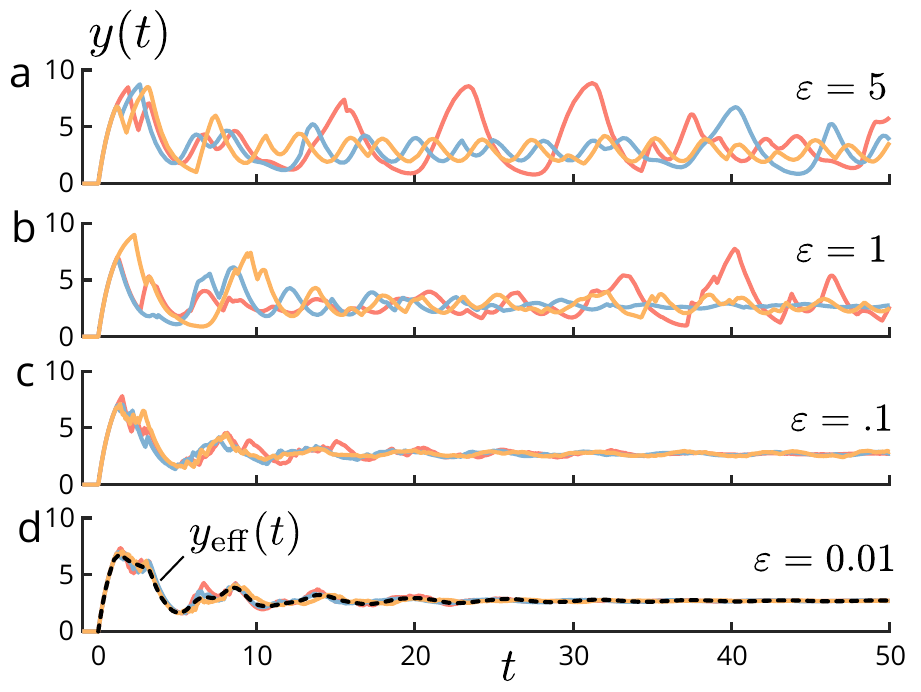}
    \caption{Stochastic realizations of the switching delayed feedback system $y(t)$ described by Eq.~\eqref{ck1} for increasingly fast switching speeds. In the top panel, slow switching ($\varepsilon=5$) yields richly stochastic solutions. In fast switching ($\varepsilon=0.1$) in the bottom panel, trajectories converge to an effective deterministic solution $y_{\mathrm{eff}}(t)$ described by Eq.~\eqref{effective}.}
    \label{fig:decreasing epsilon}
\end{figure}

We now implement stochastic switching in Eq.~\eqref{DNF1} by taking $\tau$ to evolve according to a two-state Markov process, $\tau \in \{\tau_1,\tau_2\}$ with $\mathcal{W}_{1\to 2} = \alpha$ and $\mathcal{W}_{2\to1} = \beta$ so that
\begin{equation}\label{master2}
\frac{\dd Q(\tau_1,t)}{\dd t} = \beta Q(\tau_2,t) - \alpha Q(\tau_1,t) = \beta - (\alpha + \beta)Q(\tau_1,t)
\end{equation}
The last equality follows from the fact that $Q(\tau_1,t) + Q(\tau_2,t) = 1$ for this model. When $\tau$ is fixed, Eq.~\eqref{DNF1} has a well-defined Hopf bifurcation point at $\tau = \tau_c$. Indeed, for $\tau_1, \tau_2 > \tau_c$ substituted into Eq.~\eqref{DNF1}, the system admits a limit cycle with amplitude and frequency determined by the corresponding delay. Numerically simulating Eq.~\eqref{DNF1} with switching given in Eq.~\eqref{master2} indeed shows dynamics wherein the solution jumps between limit cycles (see Figure~\ref{fig:model}b). However, in many situations, transitions between states of a PDMP are fast relative to the other dynamics of the system. Simulations of the stochastic hybrid system with $\tau_1, \tau_2 > \tau_c$ and $\alpha = \alpha/\varepsilon$ and $\beta = \beta/\varepsilon$ for $0 < \varepsilon \ll 1$ show that the system contracts to the equilibrium (see Figure~\ref{fig:decreasing epsilon}). This result is not intuitive.

\section{General Formulation}
To understand what causes the stabilization of the fixed point with fast stochastic delay switching, we consider a microscopic model for delayed negative feedback. We begin with a delayed master equation \cite{brett2014gaussian} and then invoke a van Kampen expansion to derive a Smoluchowski equation for the probability density of the stochastic variable undergoing delayed negative feedback. We begin by writing the master equation analog of Eq.~\eqref{DNF1} with $\tau$ undergoing a continuous-time discrete Markov process with $N$ states: $\{\tau_1,\tau_2,...,\tau_N\}$.

Let $d(t) \in \mathbb{N}$ be a random variable whose dynamics are governed by the reactions delineated by Eq.~\eqref{DNF1}. Thus, $d(t)$ could represent the number of translated protein molecules. Let $P(m,t,\tau)$ be the probability that $d = m$ at time $t$ and that the current delay value is $\tau$. The dynamics of $P(m,t,\tau)$ can be written as
\begin{align}
\frac{\dd P(m,t,\tau_i)}{\dd t} &= I(P(m-1,t,\tau_i) - P(m,t,\tau_i)) -\gamma(mP(m,t,\tau_i)-(m+1)P(m+1,t,\tau_i))\nonumber\\ 
&- w\left(\sum_{j=1}^{N} \sum_{M=1}^{\infty}\frac{M^n}{K^n + M^n}(P(m,t,\tau_j; M,t-\tau_i,\tau_j) - P(m+1,t,\tau_j; M,t-\tau_i,\tau_j))\right)\nonumber\\
&+ \sum_{k = 1, k\neq i}^N \mathcal{W}_{k \to i}P(m,t,\tau_k) - \mathcal{W}_{i\to k}P(m,t,\tau_i)
\label{master3}
\end{align}
The first two terms correspond to constitutive protein production and natural degradation, respectively. The transitions between distinct delay values are captured in the last two terms of Eq.~\eqref{master3}. The delayed negative feedback manifests in the middle terms of Eq.~\eqref{master3} and consists of the joint probabilities $P(m,t,\tau_j; M, t - \tau_i, \tau_j)$ of having $m$ protein molecules at time $t$ while $\tau=\tau_j$ and $M$ protein molecules at time $t-\tau_i$ while $\tau = \tau_j$. Asserting that $M$ protein molecules must exist at time $t - \tau_i$ means that delayed negative feedback can occur only if protein molecules exist to cause the negative feedback. \edit{Eq.~\eqref{master3} therefore describes a non-Markovian process since the value of $d(t)$ depends on the value of $d(t-\tau_i)$.} Thus, Eq.\eqref{master3} is not closed on the level of one-time quantities. Indeed, to determine the dynamics of the joint probability distributions, one needs to have information about three- and four-point correlations. The result is an infinite hierarchy of equations. \edit{Following a common approach \cite{chou2011non,tauber2014critical}, we assume the probability distributions for the number of protein molecules at distinct times are independent so that $P(m,t,\tau_j;M, t-\tau_i, \tau_j) = P(m,t,\tau_j)P(M,t-\tau_i,\tau_j)$. We refer the reader to the classical work of Frank \cite{frank2005delay,frank2005delayb} or the more recent \cite{loos2019fokker} for further discussion on this approximation.}

Unfortunately, little analysis can be performed on Eq.~\eqref{master3}. To make progress, we follow arguments similar to those in \cite{frank2007kramers} and invoke the van Kampen expansion by setting $x \equiv m/\mathcal{N}$ and Taylor expanding terms in Eq.~\eqref{master3} to $O(\mathcal{N}^{-1})$. \edit{Here, $\mathcal{N}$ represents the the maximum number of translated protein molecules possible, as determined by the physics of the system.} We obtain the following Chapman-Kolmogorov equations, as known for stochastic delay systems \cite{kuchler1992langevins,loos2019fokker}
\begin{align}
\frac{\partial p_i}{\partial t} =& -\frac{1}{\mathcal{N}}\frac{\partial}{\partial x}\Big((I - \gamma x)p_i(x,t)\Big) + \frac{1}{\mathcal{N}}\sum_{j=1}^N\int_0^\infty \frac{\partial p_i}{\partial x}\frac{w\xi^n}{K^n + \xi^n} p(\xi,t-\tau_i,\tau_j) d\xi\nonumber \\
&+ \sum_{k = 1, k\neq i}^N \mathcal{W}_{k \to i}p_k(x,t) - \mathcal{W}_{i\to k}p_i(x,t),
\label{ck1}
\end{align}
\edit{where $p_i(x,t) \equiv p(x,t,\tau_i)$.} The terms inside the summation and integral evaluate to the expected value of the Hill function, giving \edit{
\begin{align}
\frac{\partial p_i}{\partial t} =& -\frac{1}{\mathcal{N}}\frac{\partial}{\partial x}\left(\left(I - \gamma x - w\frac{ \chi_i^n}{K^n + \chi_i^n} \right)p_i(x,t)\right) \nonumber \\
&+ \sum_{k = 1, k\neq i}^N \mathcal{W}_{k \to i}p_k(x,t) - \mathcal{W}_{i\to k}p_i(x,t),
\label{ck2}
\end{align}}
\edit{where
$$
\mathbb{E}\left(\frac{w\xi^n}{K^n + \xi^n}\right) \equiv \frac{ \chi_i^n}{K^n + \chi_i^n}.
$$
Here, $\chi_i$ is the value of the delayed variable $\xi$ yielding the expected value of the Hill function. It is uniquely identified because the Hill function is injective. We point out that Eq.~\eqref{ck2} is a function of \emph{three} independent variables: $(x, \chi_i, t)$. Abusing notation slightly, we now set $p_i(x,t) \equiv p(x,\chi_i,t)$.}

To determine why fast switching between delays stabilizes the fixed point of Eq.~\eqref{DNF1}, we next must consider the limit where transitions between discrete delays are fast.

\subsection{Fast Switching Limit}
To consider the fast switching limit and derive the effective equation governing the dynamics of delayed negative feedback, we re-scale all transition propensities as $\mathcal{W}_{i\to j} \to \varepsilon^{-1} \mathcal{W}_{i\to j}$, with $0<\varepsilon \ll 1$. The effective equation manifests as a perturbation from the stationary measure of the Markov transition matrix governing the switching between delays. It will provide an approximation to the mean dynamics of Eq.~\eqref{ck2}. First, we rewrite Eqs.~\eqref{ck2} in matrix-vector format:
\begin{equation}
\label{ck3}
\frac{\partial }{\partial t} |p\rangle = \frac{1}{\varepsilon}\mathbf{A} |p\rangle + \mathbb{L}(|p\rangle),
\end{equation}
where 
$$
|p\rangle \equiv \left(\begin{array}{c}
p_1(x,t)\\
p_2(x,t)\\
\vdots\\
p_N(x,t)
\end{array}\right)\quad\quad \mathbb{L}(|p\rangle) \equiv \left(\begin{array}{c}
\mathbb{L}_1(p_1(x,t))\\
\mathbb{L}_2(p_2(x,t))\\
\vdots\\
\mathbb{L}_N(p_N(x,t))
\end{array}\right)
$$
and
$$
\mathbf{A} \equiv \left( \begin{array}{ccccc}
-\sum_{k\neq1}^N \mathcal{W}_{1\to k} & \mathcal{W}_{2\to 1} & \mathcal{W}_{3 \to 1} & \ldots & \mathcal{W}_{N \to 1} \\
\mathcal{W}_{1\to 2} & -\sum_{ k\neq 2}^N \mathcal{W}_{2\to k} & \mathcal{W}_{3\to 2} &\ldots & \mathcal{W}_{N\to 2}\\
\mathcal{W}_{1 \to 3} & \mathcal{W}_{2 \to 3} & -\sum_{ k\neq 3}^N \mathcal{W}_{3\to k} & \ldots & \mathcal{W}_{N \to 3}\\
 \vdots & \vdots & \vdots & \ddots & \vdots \\
 \mathcal{W}_{1 \to N} & \mathcal{W}_{2 \to N} & \mathcal{W}_{3\to N} & \ldots & -\sum_{ k\neq N}^N \mathcal{W}_{N\to k}

\end{array}\right).
$$
The linear operators $\mathbb{L}_i$ act upon any sufficiently smooth function $F$ by:
\edit{
$$
\mathbb{L}_i \edit{F} \equiv -\frac{\partial}{\partial x}\left(\left(I - \gamma x - w\frac{\chi_i^n}{K^n + \chi_i^n}\right)F\right).
$$
}
The co-kernel of $\mathbf{A}$ is spanned by the $N$-dimensional vector
$$
\langle \psi | \equiv  (1,1,\ldots,1) ,
$$
and the kernel of $\mathbf{A}$ is spanned by the $N$-dimensional vector \edit{$|\phi\rangle \equiv (\phi_1, \phi_2,...,\phi_N)^T$}, whose entries, in general, will be determined by specific relations between the transition propensities and satisfy $\langle \psi | \phi \rangle = 1$. \edit{Indeed, $|\phi\rangle$ is the invariant measure of the continuous-time Markov chain governing the jumps between delays.} 

Let $q = \langle \psi | p \rangle$ and $|w\rangle = |p\rangle - q|\phi\rangle$ so that $q$ is the component of $|p\rangle$ in the co-kernel of $\mathbf{A}$ and $|w\rangle$ is in the orthogonal complement.  Applying $\langle \psi |$ to both sides of Eq.~\eqref{ck3} gives
\begin{equation}
\label{left}
\frac{\partial q}{\partial t} = \langle \psi | \mathbb{L} | w + q\phi \rangle
\end{equation}
Substituting $|p\rangle = |w + q\phi\rangle$ into Eq.~\eqref{ck3}  and using the fact that $|\phi\rangle$ is in the kernel of $\mathbf{A}$, we obtain
\begin{equation}
\label{mid}
\frac{\partial }{\partial t} |w\rangle = \frac{1}{\varepsilon} \mathbf{A}|w\rangle + \left(\mathbb{I}_N - |\phi\rangle\langle \psi| \right)\mathbb{L}(|w + q\phi\rangle),
\end{equation}
where $\mathbb{I}_N$ is the $N \times N$ identity matrix. Introduce the expansion
$$
|w\rangle = |w_0\rangle + \varepsilon |w_1\rangle + O(\varepsilon^2)
$$
and substitute into Eq.~\eqref{mid}.  Collecting $O(\varepsilon^{-1})$ terms gives $\mathbf{A}|w_0\rangle = |0\rangle \Rightarrow |w_0\rangle = |0\rangle$. Because we are only interested in the mean dynamics and not the fluctuations about the mean, we need not attempt to calculate the higher-order terms. We hope to explore this in future work. 

Substituting the leading order term for $|w\rangle$ into Eq.~\eqref{left} gives the Smoluchowski equation for the probability density $q(x,\pmb{\chi},t)$ for a protein undergoing delayed negative feedback at time $t$: 
\edit{\begin{equation}
\label{smoluchowski1}
 \frac{\partial q}{\partial t} = \langle \psi|\mathbb{L}|q\phi\rangle = \sum_{n=1}^{N}\phi_n \mathbb{L}_n q(x,\pmb{\chi},t),
\end{equation}}

\noindent\edit{where $\pmb{\chi} \equiv (\chi_1,\chi_2,...,\chi_N)$. Here, $q(x,\pmb{\chi},t)\Delta x = \mathbb{P}(y(t) \in (x,x+\Delta x)) + O(\Delta x^2)$ and $q(x,\pmb{\chi},t)\Delta \chi_i = \mathbb{P}(y(t-\tau_i) \in (\chi_i,\chi_i+\Delta \chi_i)) + O(\Delta \chi_i^2)$. In words, Eq.~\eqref{smoluchowski1} says that the leading-order effective dynamics evolve via the weighted average of the different subsystems, where the weight is determined by the stationary distribution of the underlying Markov chain controlling switching. An important caveat appears: that such a stationary distribution exists. For all examples within the realm of models of genetic feedback, we anticipate this to be the case, but the breakdown of this assumption may be of future mathematical interest. }

\edit{Another 
 area of mathematical interest is that  Eq.~\eqref{smoluchowski1} describes the dynamics of $q(x,\pmb{\chi},t$) with a multivariate Smoluchowski equation, therefore assuming that the current state of the random variable is \emph{independent} of its value at all previous times. It is a \textit{Markovian} description of a \emph{non-Markovian} process---the fact that this is a delayed negative feedback system necessarily renders it non-Markovian.  We find that Eq.~\eqref{smoluchowski1} is a good approximation to the non-Markovian process in the fast-switching limit (see Fig.~\ref{fig:decreasing epsilon}d.). We take this as evidence that the independence assumption we made works well in the fast switching limit. We hope to quantify this approximation and precisely determine where it breaks down in future work.}

\edit{One notable feature of the structure of the effective dynamics \eqref{smoluchowski1} is that the system cannot be described by a single effective delay. Rather, the leading-order dynamics evolve by the simultaneous influence of all all delay subsystems in the fast-switching limit.}

\section{Delayed Negative Feedback Again}
We now invoke Eq.~\eqref{smoluchowski1} and apply it to the specific example discussed in Section ~\ref{sdnf}. In this case, the resulting Chapman-Kolmogorov equations are
\edit{
\begin{align*}
\frac{\partial p_1}{\partial t} =& -\frac{\partial}{\partial x}\left(\left(I - \gamma x - w\frac{\chi_1^m}{K^m + \chi_1^m}\right)p_1(x,t)\right) \\
&- \alpha p_1(x,t) + \beta p_2(x,t)\nonumber\\
\frac{\partial p_2}{\partial t} = & -\frac{\partial}{\partial x}\left(\left(I - \gamma x - w\frac{\chi_2^m}{K^m + \chi_2^m}\right)p_2(x,t)\right) \\
&+ \alpha p_1(x,t) - \beta p_2(x,t)\nonumber
\end{align*}
}
Here, $\langle \psi | = (1,1)$ and 
$$
|\phi \rangle = \frac{1}{\alpha + \beta}\left ( \begin{array}{c}
\beta\\
\alpha
\end{array} \right ),
$$
so that the resulting Smoluchowski equation is
\begin{align}
\frac{\partial q}{\partial t} = &-\frac{\beta}{\beta + \alpha}\frac{\partial}{\partial x}\left(\left(I - \gamma x - w\frac{\chi_1^n}{K^n + \chi_1^n}\right)q(x,\pmb{\chi},t)\right)\nonumber\\
&-\frac{\alpha}{\beta + \alpha}\frac{\partial}{\partial x}\left(\left(I - \gamma x - w\frac{\chi_2^n}{K^n + \chi_2^n}\right)q(x,\pmb{\chi},t)\right).
\end{align}
Thus, the mean dynamics of the protein undergoing delayed negative feedback is given by the kinetic equation
\begin{equation}
\label{effective}
\frac{\dd y_{\rm{eff}}}{\dd t} = I - \gamma y_{\rm{eff}} - \frac{\beta}{\beta + \alpha}\frac{ w y_{\rm{eff}}(t-\tau_1)^n}{K^n + y_{\rm{eff}}(t-\tau_1)^n} - \frac{\alpha}{\beta + \alpha}\frac{ w y_{\rm{eff}}(t-\tau_2)^n}{K^n + y_{\rm{eff}}(t-\tau_2)^n}.
\end{equation}
\edit{The stability of this averaged two-delay system can surprisingly display behavior distinct from either of the two subsystems, as we show in the next subsection.}

\subsection{Stabilization via Stochastic Switching}
Setting the time derivative equal to zero in Eq.~\eqref{effective} and solving for the equilibrium gives the same solution $y^*$ as obtained from Eq.~\eqref{equilibrium}. The linearization of Eq.~\eqref{effective} about $y^*$ yields
$$
\frac{\dd u}{\dd t} = -\gamma u - \frac{w\beta}{\alpha + \beta}f'(y^*)u(t-\tau_1) - \frac{w\alpha}{\alpha + \beta}f'(y^*)u(t-\tau_2)
$$
where $u(t) \equiv y_{\rm{eff}}(t) - y^* $.
\begin{figure}[ht]
    \centering
    \includegraphics[width=3in]{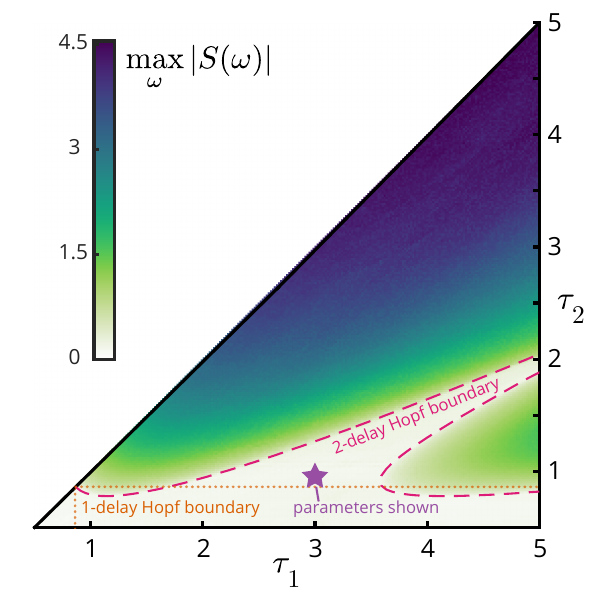}
    \caption{Bifurcation structure for the effective delay equation found in \eqref{effective}. In the heatmap, the maximum of the power spectrum $S(\omega)$ is shown. The dashed boundary lines for the 2-delay system correspond to the linear stability analysis in Eq.~\eqref{hopf2}. The dotted lines correspond to the Hopf boundary for a single delay Eq.~\eqref{hopf1}. The star shows the choice of delays used elsewhere unless noted otherwise. For these delays, the oscillations of each subsystem are stabilized by sufficiently fast switching. Symmetric transitions are considered $\alpha=1,\beta=1$ so only $\tau_1\leq \tau_2$ need be considered. }
    \label{fig:hopf_2delay}
\end{figure}

\noindent As in Section~\ref{hopf}, we invoke the ansatz $u(t) = Ae^{\lambda t}$ and determine $\lambda$ from the auxiliary equation
$$
\lambda = -\gamma - \frac{w\beta}{\alpha + \beta} f'(y^*) e^{-\lambda \tau_1}- \frac{w\alpha}{\alpha + \beta} f'(y^*) e^{-\lambda \tau_2}
$$
Setting $\lambda = i\omega$ yields the following conditions for a Hopf bifurcation in Eq.~\eqref{effective}:
\begin{align}
\begin{split}\label{hopf2}
\omega &= w \frac{\alpha}{\alpha + \beta}f'(y^*)\sin{(\omega\tau_1)} + w \frac{\beta}{\alpha + \beta}f'(y^*)\sin{(\omega\tau_2)}\\
-\gamma &= w \frac{\alpha}{\alpha + \beta}f'(y^*)\cos{(\omega\tau_1)} + w \frac{\beta}{\alpha + \beta}f'(y^*)\cos{(\omega\tau_2)}\\
-\frac{\omega}{\gamma} &= \frac{\alpha \sin{(\omega \tau_1)} + \beta \sin{(\omega \tau_2)}}{\alpha \cos{(\omega \tau_1)} + \beta \cos{(\omega \tau_2)}}
\end{split}
\end{align}
In Figure~\ref{fig:hopf_2delay}, we show the locus of Hopf bifurcation points in parameter space for Eq.~\eqref{effective} and compare it to the Hopf bifurcation points for the single delay equation given in Eq.~\eqref{DNF1}. We can see that there are regions of parameter space wherein $\tau_1, \tau_2$ are larger than the single delay critical Hopf value but the system continues to reach the fixed point. Hence, the fast switching between delays in the stochastic system causes the effective behavior of the system to behave as if the feedback followed two distinct delay values simultaneously. The presence of multiple delays increased the range of delay values for which the fixed point was stable. 

\begin{figure}
    \centering
    \includegraphics[width=\textwidth]{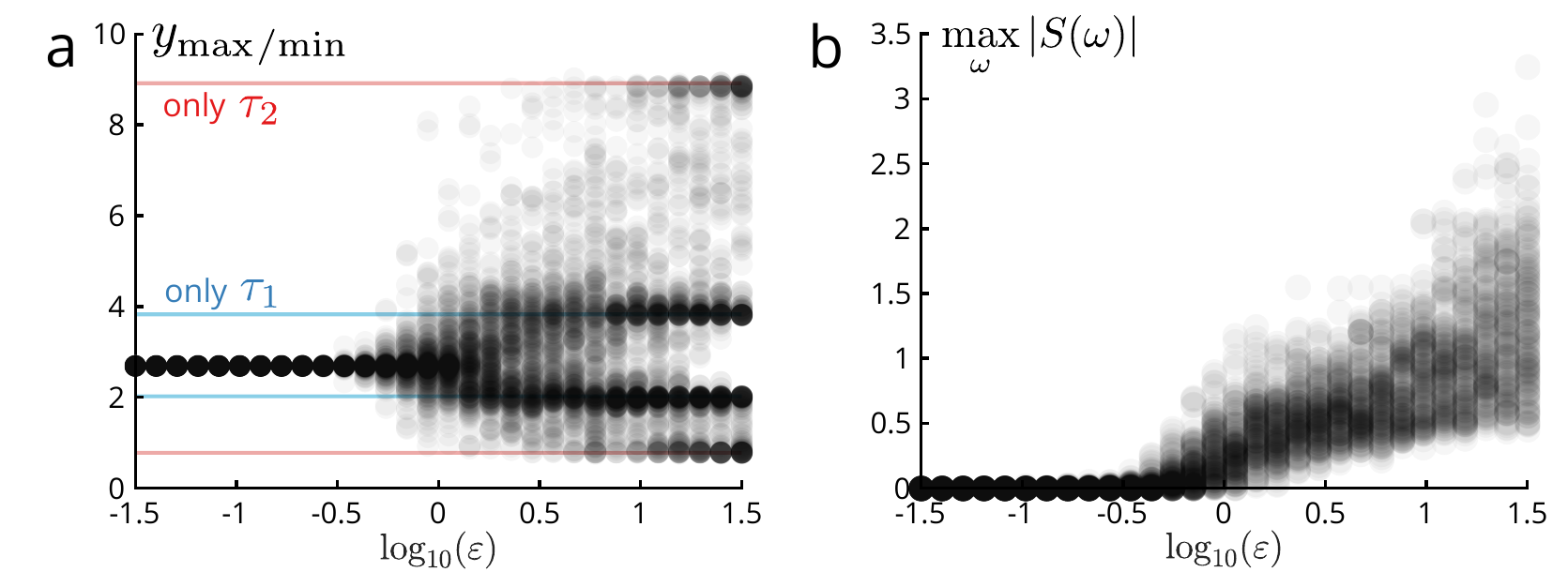}
    \caption{Bifurcation structure of the stochastically switching delay system as a function of the overall switching rate $\varepsilon$. Each circle is a stochastic realization, with 500 total per parameter set. \textbf{a}: The maximum and minimum values over a time window in stochastic simulation show that as switching gets slower, oscillations emerge. Solid lines represent the oscillation peaks for the single delay subsystems with corresponding delays. \textbf{b}: The peak of the power spectrum $\max_\omega|S(\omega)|$ also demonstrates a  sub-critical Hopf bifurcation parameterized by the switching rate $\varepsilon$.}
    \label{fig:bifur_eps}
\end{figure}

When switching is not fast, then the increased stabilization of the fixed point is not observed. Although it is challenging to analytically investigate this scenario, numerical investigation via stochastic simulation is straightforward and can be seen in Figure \ref{fig:bifur_eps}. Explicit stochastic simulations are performed by sampling the continuous-time Markov chain and solving the delay differential equation between these events. The system is simulated for $t\in[0,100]$, and over the window $t\in[90,100]$, the minimum and maximum values are taken, as presumed magnitudes of any oscillations after the transient portions have decayed. As waiting times are increased ($\varepsilon$ large), the system spends enough time in each delay state so that the effective dynamics follow a single delay equation for the duration of time in that state. Periodic solutions corresponding to the delay of the state emerge. As the waiting time is decreased ($\varepsilon$ small), the effects of the second delay term emerge and the stabilization of the fixed point is observed (see Figures ~\ref{fig:decreasing epsilon} and \ref{fig:bifur_eps}).

\section{Conclusion}

We summarize the main contributions of the manuscript as follows. Most generally, we have derived an effective delay equation in the limit of fast switching between subsystems with different delays that evolve via a continuous-time Markov chain. \textit{A priori}, it is not clear whether  the behavior when rapidly switching between  systems can be replaced by one effective delay. Here, we answer that possibility in the negative for non-linear systems, similar to the semi-discretized case \cite{sadeghpour2018can}. We used this result to investigate a classical model of delayed negative feedback with a new twist of stochastic switching between two delays. In our stochastic model, we showed that sufficiently fast stochastic switching between two delays stabilizes the system where each delay alone produces oscillations. 

\edit{Our results sit within broader biological and mathematical contexts. First, we note the relation to the literature on distributed delay systems, especially in models of genetic feedback. The effective dynamics derived here arising in the fast switching limit \eqref{effective} are exactly the form of distributed delay descriptions of genetic feedback considered elsewhere \cite{bernard2009distributed,josic2011stochastic,brett2013stochastic}. We have therefore provided further mechanistic motivation for the inclusion of these distributed delay systems. We show that a Hopf bifurcation in the total switching rate occurs, indicating that fast switching and slow have fundamentally different behavior. This nuance in timescales of stochasticity does not exist in descriptions with distributed delays.} 

Stochastically switching delays add a new vignette to the broader theme of stochasticity in genetic feedback.  Importantly, we consider stochasticity \textit{only} in the delay to emphasize its impact on the behavior of the system. This is in contrast with other studies where stochasticity is included in genetic feedback in other ways and new behaviors appear. For instance, molecule counts in the genetic machinery are low enough to justify exploring demographic fluctuations \cite{galla2009intrinsic}. This randomness is known to destabilize oscillations \cite{potoyan2014dephasing}. It is therefore of  future interest to investigate how stochastic switching of delays and demographic fluctuations intertwine. A natural starting point would be similar investigations for non-delay systems, \cite{hufton2016intrinsic}, but we anticipate further challenges due to the hierarchy of multi-point correlations for stochastic delay systems as discussed in \cite{frank2005delay,frank2005delayb,frank2007kramers,loos2019fokker}. Other studies also include stochastic switching with delays arising from binding and unbinding of promoters \cite{zambrano2015interplay,cao2020analytical} producing bursting behavior. These models have rich analytical tractability but do not include different delays. It remains to be explored whether these calculations can be extended to different delays, as considered here. To further the biological relevance, it is also pressing to develop a more mechanistic explanation of how stochastic switching of delays may arise. One intriguing direction is the emergence of stochastic switching from  dual delay feedback pathways, such as in NF-$\kappa$B signaling\cite{venturelli2012synergistic,longo2013dual}.

\edit{Lastly, on the purely mathematical side, our model and analysis add to the mosaic of stochastic systems that behave fundamentally differently than their deterministic counterparts or subsystems. Specifically, our results first add to a long history along the theme of how noise can stabilize systems \cite{arnold1983stabilization,mackey1990noise,basak2001stabilization,bernard2009distributed,gupta2013transcriptional}. Secondly, they provide another example of how stochastic switching can result in unexpected behavior of stable solutions to the corresponding non-switching ODEs and PDEs~\cite{benaim2014stability,lawley2014sensitivity,benaim2016lotka,lawley2018blowup}}. Although it may seem restrictive that we consider a Markov chain model by which the delays evolve, an arbitrary delay distribution may be constructed via the theory of phase-type distributions \cite{commault2003phase}. We were unable to compute analytical results when switching was not fast, or even a next-order correction to the leading-order behavior. It is perhaps feasible to use a moment-based approach that others have used in stochastically switching delayed \cite{verriest2009stability,bernard2015,gomez2016stability,sykora2020moment} or other stochastic hybrid systems \cite{lawley2015stochastic,bressloff2015moment}. Alternatively, it may be feasible and interesting to investigate the opposite limit of the one considered, where  the dynamics of the subsystems are fast relative to the switching, as seen in many other biological systems \cite{miles2018analysis,marbach2022coarse} and is perhaps readily handled by classical homogenization techniques \cite{pavliotis2008multiscale}. Further, the dynamical systems behavior of the derived effective \eqref{effective} system could also be probed more thoroughly, perhaps using the methods of \cite{du2019two} that compute normal forms and investigate higher co-dimension bifurcations.

\section*{Acknowledgements}
BK and CM would like to thank Mehdi Sadeghpour for insightful discussions. BK would like to thank his son, Surya, for "delaying" his arrival enough to complete this manuscript.

\bibliography{references-delays} %

\end{document}